\documentclass[%
 reprint,
 amsmath,amssymb,
 aps,
]{revtex4-2}

\usepackage{graphicx}
\usepackage{dcolumn}
\usepackage{bm}
\usepackage{xcolor}
\usepackage{booktabs}
\usepackage{amsmath}

\begin{document}

\preprint{APS/123-QED}

\title {Multiscale Phase Separation in Chemophoretic Active Matter}

\author{Manisha Jhajhria$^{1}$, Subir K. Das$^{2\,*}$, and Snigdha Thakur$^{1}$}
\email{das@jncasr.ac.in}
\email{sthakur@iiserb.ac.in}

\affiliation{$^{1}$Department of Physics, Indian Institute of Science Education and Research Bhopal, Madhya Pradesh 462066, India\\
$^{2}$Theoretical Sciences Unit, Jawaharlal  Nehru Centre for Advanced Scientific Research, Jakkur P.O., Bangalore 560064, India}

\begin{abstract}

Nonreciprocal interactions in active matter provide interesting structure and dynamics. Here we investigate chemophoretic systems in which nonreciprocity arises from the asymmetric coupling between agents: first species produces certain chemicals and the other phoretically responds to it. This leads to phase separation at varying scales. Our study uncovers a re-entrant steady-state phase diagram as the nature of the coupling changes from chemoattractive to chemorepulsive character. Chemoattraction provides sustained domain growth, leading to macrophase separation via cluster coalescence. Aggregation in the chemorepulsive case, on the other hand, leads to a steady-state situation that displays phase separation only at a microscale, owing to strong caging effect and frequent fragmentation. The overall far-from-steady-state dynamics is quantified via calculations of growth exponents, cluster transition matrices, and mean-squared displacements.

\end{abstract}

\maketitle
\section{\label{sec:level1}Introduction}

Active matter systems consist of self-driven units that continuously draw energy from the surroundings and often convert it into directed motion, thereby operating far from equilibrium~\cite{sriram2010,thomas2001, marchetti2013,cavagna2014,bechinger2016active, gompper20202020,vicsek2012collective,Zhang2010,sanchez2012spontaneous}. Examples range from bacterial suspensions and cytoskeletal assemblies in the pure biological domain to synthetic colloids and other chemically powered assemblies of particles~\cite{Zhang2010,sanchez2012spontaneous,jiang10,Howse2007, wang2015one,theurkauff2012,BIALKE2015}. A few defining features of active matter are nonreciprocal interactions and emergence of collective phenomena, even in the absence of interparticle potential, having no equilibrium counterparts. Examples include~\cite{buttinoni2013, marchetti2013,bechinger2016active,toner1998flocks,gompper20202020,Caporusso2023,wysocki2016propagating} motility-induced phase separation, breaking of continuous symmetry at low dimensions, etc. Understanding related \textit{nonequilibrium} phase behavior and corresponding transition kinetics remains a central challenge in the field~\cite{vicsek1995,Toner1995,Czirok1997,thakur12, Redner2013,fily2012,stehammar2013,midya2017,annrev2015, BIALKE2015,caprini2020,binder2021phase,ben2016, benjamin2017,gabriel2013,siebert2018critical,das2017pattern,saha2014clusters,stenhammar2015,dittrich2023growth,mj2025}.

Among various possibilities that may generate nonreciprocity, chemophoretic interactions are of utmost importance~\cite{theurkauff2012,sturmer2019chemotaxis,dhruv2017,golestanian2019}. Chemically active particles can produce concentration gradients of chemicals to which certain constituents may respond. This may lead to effective interactions among particles which are of uneven nature~\cite{Golestanian2007,golestanian2019,theurkauff2012,dhruv2017, C8CC06467A, mj2023,sturmer2019chemotaxis,saha2014clusters,stark2018artificial,liebchenbeno2018,prabha2018,soto2014self,Ivlev2015,liebchenbeno2018, Saha2019,golestanian2019,grauer2020swarm,dinelli2023non}. Such nonreciprocity can generate a plethora of non-conventional situations, including traveling clusters, self-organized currents, and varieties of pattern-forming instabilities~\cite{cates2010arrested, dhruv2017, kryuchkov2018dissipative, golestanian2019, schmidt2019light, grauer2020swarm, Fruchart2021,namita2022,Sandeep2023}. While recent works have significantly advanced our understanding of steady-state organizations and phase diagrams of these systems, little attention has been paid to related \emph{kinetic pathways} through which phases emerge. In the latter context, important directions are related to the studies of the dynamical routes through which density fluctuations grow, clusters nucleate, and domains coarsen~\cite{vicsek1995, bray2002, baglietto2012criticality, Peruani_2013, Cates2015,  chakraborty2020relaxation, paul2021clusters, Das2022, dittrich2023growth, paul2024finite, mj2025}. Here our focus is on phase behavior and evolution kinetics in one such model system.

We consider binary mixtures~\cite{mj2023,mj2025,stark2018artificial} of chemically active and phoretic particles. While chemically active particles act as sources or sinks of a chemical field, the phoretic particles respond to the resulting gradients by migrating either toward or away from the latter. Depending on the signs and strengths of chemical activity and phoretic mobility, the effective interactions between the two species can be attractive or repulsive~\cite{dhruv2017, mj2025, sturmer2019chemotaxis, schmidt2019light, dinelli2023non}. 
Experiments with catalytic Janus swimmers interacting with passive tracers have revealed long-range chemophoretic attraction, clustering, and directed transport induced solely by chemical gradients \cite{dhruv2017, wang2019interactions, katuri2021inferring, singh2022interaction, huang2020inverse, hauke2020clustering, gao2017dynamic, mu2022binary}. 
In contrast, the opposite tendency of motion away from chemical generating source has received comparatively less attention. Nevertheless, a few recent experiments on catalytic colloids interacting with passive tracers have reported systematic exclusion zones and outward tracer fluxes arising from diffusiophoretic repulsion~\cite{wang2018visible, huang2020anisotropic}. 
These motivate theoretical exploration of understanding how the kinetics of phase separation in chemorepulsive couplings differ from those in chemoattractive couplings in binary mixtures~\cite{liebchen2015}.

A central question that follows is how the nature and strength of phoretic interactions influence the scaling properties of structure formation and domain growth as the systems evolve toward steady states. In passive phase-ordering kinetics, coarsening is often well described by established dynamic scaling, in which growth laws are influenced by factors such as conservation laws, order-parameter symmetry, and dimensionality~\cite{bray2002,lifshitz1961kinetics,binder1974theory,binder1977theory,siggia1979late}. For nonequilibrium transitions with intrinsically nonreciprocal interactions, however, the validity of such scaling arguments is not guaranteed. Here we address associated questions by systematically examining the phase-ordering kinetics of a chemophoretic binary mixture and directly comparing results from chemoattractive and chemorepulsive couplings. 

We study both chemoattractive and chemorepulsive cases, under a common framework, via incorporation of Langevin dynamics. Although phase separation is observed in both the situations, our results reveal fundamentally distinct nonequilibrium pathways and steady-state features in the two cases. In the chemoattractive regime, growth occurs via cluster-cluster coalescence, that leads to rapid increase in domain size and thus, macrophase separation. In contrast, chemorepulsive couplings give rise to aggregation that is driven by activity-induced kicks. Furthermore, growth in the latter case is arrested, owing to strong caging and fragmentation. As a consequence, the system stabilizes microphase-separated steady states composed of finite-sized, dynamically fluctuating domains. These qualitative differences between the two cases are reflected consistently across the steady-state phase diagram, coarsening exponents, a transition-matrix analysis, and single-particle dynamics. Our results show that chemophoretic interactions fundamentally reshape phase-ordering kinetics and place the observed growth laws outside equilibrium universality.

\section{\label{sec:model}Models and Techniques}
Our model systems consist of two-dimensional binary suspensions of chemically active source ($S$) and chemically inactive phoretic ($P$) particles. While each $S$ particle generates a radially-symmetric chemical field around it, the $P$ particles sense the chemical gradient around them. The latter species responds to the chemical signal via a diffusiophoresis mechanism by moving towards (chemoattraction) or away (chemorepulsion) from the source particles. A system contains total $N~(=N_S+N_P)$ colloids, with $N_S$ and $N_P$ being the numbers of source and phoretic colloids, respectively. We have fixed the fraction of source, $N_S/N$, at $0.2$, same as our previous study~\cite{mj2023,mj2025}.

The motion of each colloid is governed by Langevin dynamics. Related dynamical equation for particle $i$, at a temperature $T$, is given by 
\begin{equation}\label{lgv}
m{{\ddot{\bf{r}}_{i}}} = {{\bf{F}}_i} - \gamma \dot{\bf{r}}_i
+\sqrt{2\gamma k_{B}T}{\boldsymbol{\xi}_{i}}.
\end{equation}
Here $m$ is the mass of each colloid, $\gamma$ is a friction coefficient, and $\boldsymbol{\xi}_i$ represents Gaussian white noise with zero mean and unit variance. The components of $\boldsymbol{\xi}_i$ satisfy
$\langle \xi_{i\alpha}(t)\,\xi_{j\beta}(t') \rangle 
=\delta_{ij}\,\delta_{\alpha\beta}\,\delta(t-t')$, for two different times $t$ and $t'$,
where $\alpha$ and $\beta$ denote Cartesian indices~\cite{frenkel2023understanding}. In Eq.~(\ref{lgv}), ${{\bf{F}}_i}$ is the net force acting on each colloid, which depends on the type of the colloid. The $S$ particles experience only excluded-volume interactions modeled via the Weeks-Chandler-Anderson (WCA) potential~\cite{wca1971}: 
\begin{equation}\label{eq:nonneighpotential}
U_{WCA}(r_{ij}) = 4\epsilon \Bigg[\Big(\frac{\sigma}{r_{ij}}\Big)^{12}-\Big(\frac{\sigma}{r_{ij}}\Big)^{6} + \frac{1}{4}\Bigg] \Theta(r_{ij}- r_{c}). 
\end{equation}
Here $\Theta$ is the Heaviside function, $\sigma$ is the diameter of each colloid, $r_{ij}=|{{\bf{r}}_i-{\bf{r}}_j}|$ and $r_c$ $(=2^{1/6}\sigma)$ is the potential cut-off radius. On the other hand, the $P$ particles experience an extra chemophoretic force $\textbf{F}_{c,i}$,  due to the source-generated chemical gradient, in addition to the WCA interaction. Therefore, the net force~\cite{sturmer2019chemotaxis,mj2025,jaiswal2024diffusiophoretic} on $P$ is $\textbf{F}_{i}^{P}(\textbf{r}_i) = -\boldsymbol{\nabla}U_{WCA}(r_i)+ \textbf{F}_{c,i}(\textbf{r}_i)$. The functional form of $\textbf{F}_{c,i}$ is given by~\cite{sturmer2019chemotaxis,roca2022self}
\begin{equation}\label{eq:chemforce}
    \mathbf{F}_{c,i}({\bf r}_{i}) = -k\boldsymbol{\nabla} c({\bf r}_{i}),
\end{equation}
where $\boldsymbol{\nabla}  c({\bf r}_{i})$ is the net chemical concentration gradient felt by a phoretic colloid, due to all source particles. The instantaneous static chemical concentration field $c(r)$, experienced by a phoretic particle, has been modeled as~\cite{mj2025}
\begin{equation}{\label{eq2}}
    c(r) = c_0 -\frac{k_0}{4\pi D_c} \sum_{j \in S} \frac{1}{|\textbf{r} - \textbf{r}_j|^2},
\end{equation} 
where $c_0$ is the reference background fuel that is consumed by each $S$, $k_0$ is the reaction rate, a higher $k_0$ implying faster fuel consumption, and $D_c$ is the chemical diffusivity. For simplicity, we have fixed the value of $k_0/4\pi D_c$ at unity. 

The most important working parameter of the model is $k$, of Eq.~(\ref{eq:chemforce}), that controls the phoretic strength: $k>0$ models chemoattraction, meaning all phoretic colloids get attracted towards source particles, forming aggregates; $k<0$ implies chemorepulsion, wherein the phoretic particles move away from the source species. Interestingly, both chemoattraction and chemorepulsion lead to phase separation. However, their phase separation kinetics is drastically different. To probe the kinetics and determine the coarsening mechanisms we computed several relevant quantities. 

The structural aspects of domains are quantified using the two-point equal time correlation function, $C(r,t)$, defined as~\cite{bray2002} 
\begin{equation}\label{eq:1}
    C(r,t) = \langle \psi (\textbf{r},t) \psi (\textbf{0},t) \rangle - \langle \psi (\textbf{r},t) \rangle \langle \psi (\textbf{0},t) \rangle.
\end{equation}
Here $\psi(\textbf{r},t)$ is a space ($\textbf{r}$)- and time (t)-dependent order parameter. For calculating $C(r,t)$, the continuum systems are first mapped onto those with an underlying two-dimensional square lattice. In this method each colloid is moved to its nearest lattice site and finally, depending upon whether a site has a particle or not, it is assigned an order parameter value $\psi(\textbf{r},t)=+1$ or $-1$, respectively~\cite{majumder2011diffusive,sutpata2013,paul2024finite}.  Furthermore, to get a pure domain morphology, a certain level of coarse-graining is applied, that eliminates the noise. This is done by updating the sign of $\psi(\textbf{r},t)$ at each site $i$ by the sign of the majority of its neighbors lying within a circle of cutoff distance $r_c'(=3.6\sigma)$.

In an evolving system, following a quench to the coexistence region, the growth is typically self-similar. Consequently, $C(r,t)$ obeys the dynamic scaling~\cite{bray2002,puri2009kinetics,onuki2002phase}
\begin{equation}\label{eq:scaling}
C(r,t) \equiv \tilde{C}\left(\frac{r}{\ell(t)}\right),
\end{equation}
where $\tilde{C}$ is a time-independent master function. The average domain length, $\ell(t)$, follows a power-law behavior~\cite{bray2002},
\begin{equation}\label{eq-scale}
\ell(t) \propto t^{\alpha},
\end{equation}
$\alpha$ being the growth exponent. In passive mixtures, if coarsening is governed by particle diffusion, one obtains $\alpha = 1/3$, known as the Lifshitz-Slyozov (LS) exponent~\cite{bray2002,lifshitz1961kinetics}. In contrast, if coarsening is driven by droplet diffusion and coalescence, then the expected growth law is $\alpha = 1/d$, $d$ being the spatial dimension, The latter is often referred to as the Binder-Stauffer (BS) exponent~\cite{binder1974theory,binder1977theory,siggia1979late,sutpata2013,midya2017droplet}. However, for nonequilibrium phase transitions, associated scaling picture may deviate significantly from the passive expectations~\cite{cremer2014scaling,das2017pattern,Linden2019,mj2025}.

All simulation parameters are reported in dimensionless units of mass $m$, length $\sigma$ and energy $\epsilon$. The unit of time is expressed as $\tau=\sqrt{m\sigma^2/\epsilon}$. We have fixed the temperature at $k_BT=0.2\epsilon$, unless stated otherwise. The equation of motion is integrated using the Verlet velocity scheme with a time step of size $\Delta t=10^{-3}\tau$. The lateral system dimension $L$ and packing fraction are taken as $150\sigma$ and $\phi=0.1$, respectively, if not stated otherwise. Periodic boundary conditions are applied in both directions. The phoretic parameter $k$ is varied between $-50\epsilon$ to $50\epsilon$ in order to probe chemorepulsive ($k<0$) as well as chemoattractive ($k>0$) regimes.

\section{Results}
\begin{figure}[ht!]
    \centering
   \includegraphics[width=\linewidth]{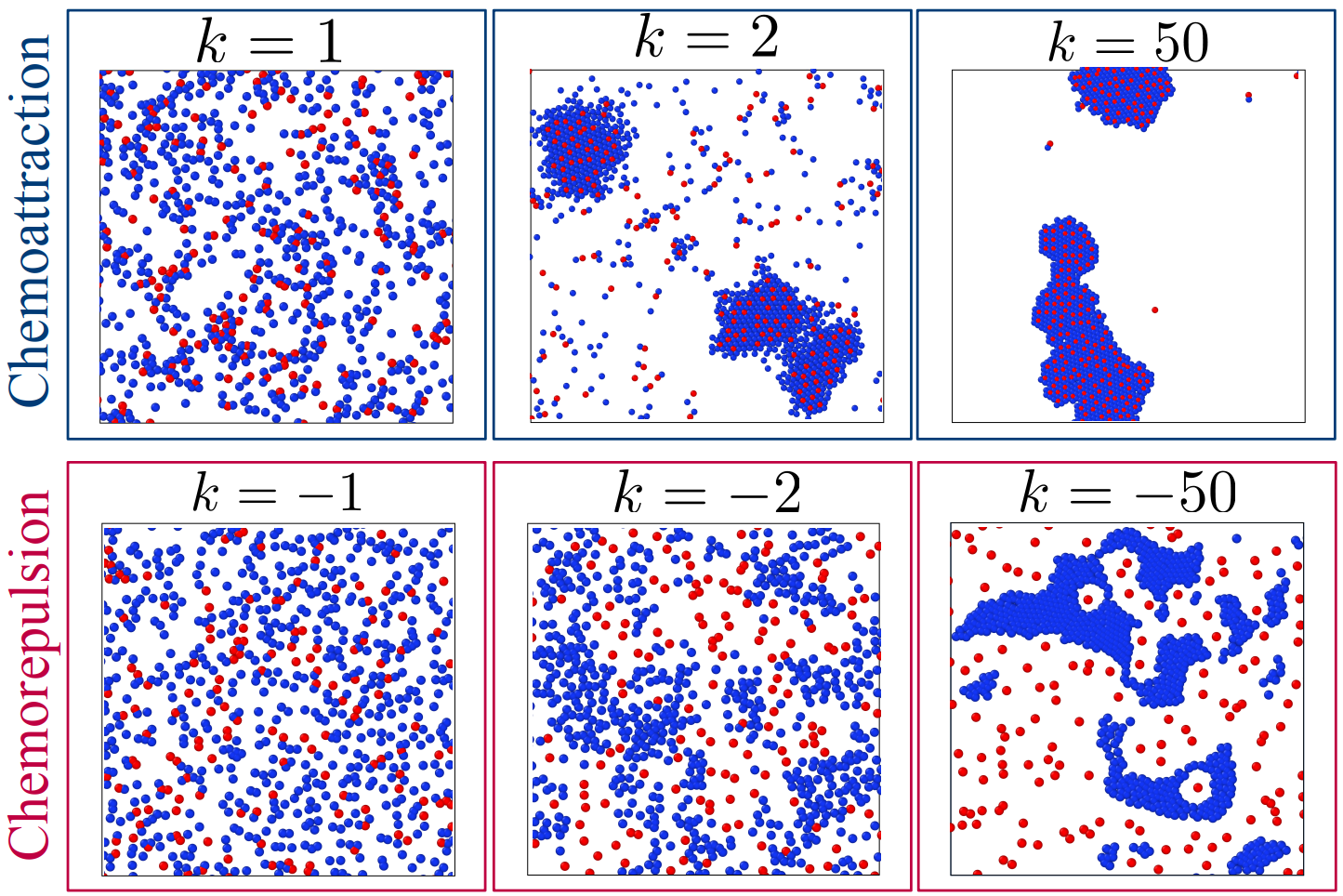}
    \caption{Instantaneous late time ($t=2000$) simulation snapshots showing aggregation in binary mixtures of source (red) and phoretic (blue) particles. The top panels display chemoattractive cases with clusters containing both $S$ and $P$ particles. The bottom panels show cases for chemorepulsion, where the clusters consist solely of $P$ particles. The diffusiophoretic strength in both cases has been kept at $|k|= 1,~2,~50$. Only $(L/4) \times (L/4)$ parts of the original system $(L=150)$ are shown for clarity.}   
    \label{fig:snap}
\end{figure}
To study the phase-separation kinetics, we quench the homogeneous mixtures of $S$ and $P$, prepared at $k=0$, to state points with non-zero $k$. We observed that, depending on the sign of $k$, the phoretic particles get attracted or repelled from the source and form particle-rich and particle-poor domains for sufficiently large magnitudes of $k$. Fig.~\ref{fig:snap} shows the late time simulation snapshots for both chemoattraction and chemorepulsion cases ($S$: red and $P$: blue). In the chemoattractive mixtures (top panels), the source particles attract the phoretic particles, leading to aggregates containing both $S$ and $P$ particles. We observed that larger values of $k$ lead to stronger self-assembly. On the other hand, in the chemorepulsive case (bottom panels), the aggregates form because of $P$ getting repelled by $S$ from all directions. Thus, the aggregates consist of only $P$ particles. The nature of these aggregates differs not only in terms of constituents, the shapes are also different from the chemoattractive case. While fractal-like domains emerge in the repulsive case, dense compact clusters appear in the chemoattractive case. In this paper, we aim to quantify this interesting difference between the two cases as well.
\begin{figure}[ht!]
    \centering
    \includegraphics[width=1.0\linewidth]{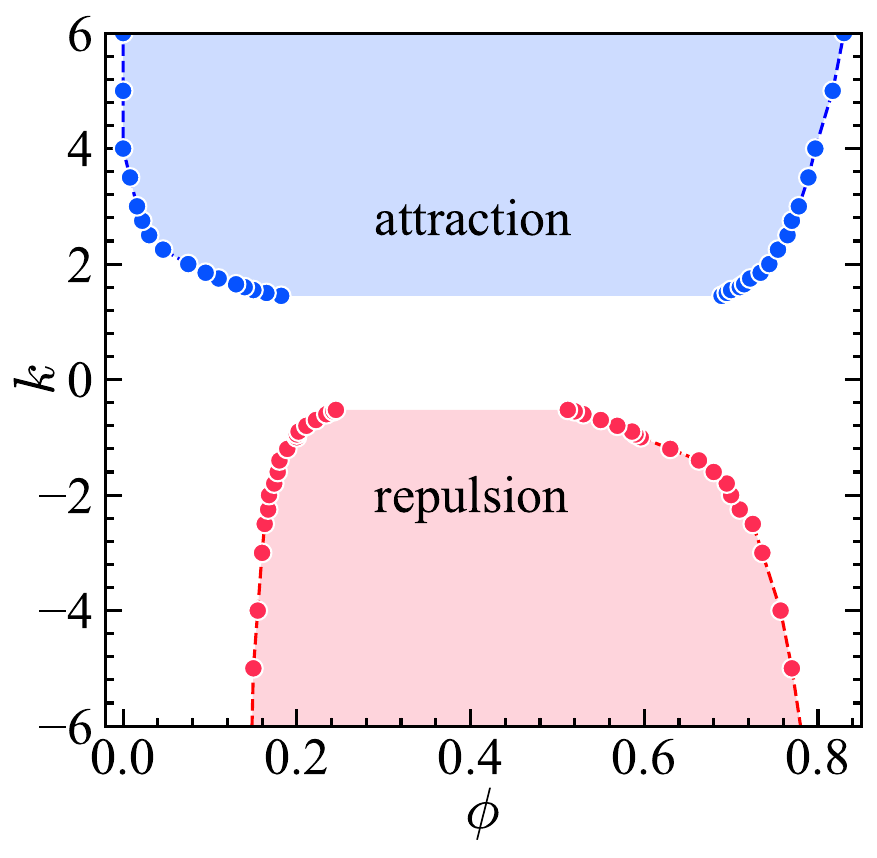}
    \caption{Steady-state coexistence curves for the binary suspensions in the phoretic strength ($k$) vs local packing fraction ($\phi$) plane. For $k<0$, the system exhibits a phase separation due to chemorepulsion (red), and for $k>0$, the system models chemoattraction (blue).}
    \label{fig:phasediag}
\end{figure}

Before discussing the cluster growth dynamics, we describe a steady-state phase diagram that is shown in Fig.~\ref{fig:phasediag}. The diagram is obtained by quenching systems of a fixed overall packing fraction $\phi=0.4$ and composition as mentioned in the previous section, from an initially spatially uncorrelated state located at $k=0$, to finite values of $|k|$. For ranges of $k$, the steady-state local density distributions develop pronounced bimodal structures, indicating phase separation~\cite{Binder_1987,Yeomans1992}. The locations of the density peaks are used to determine the coexistence points, loci of which define the phase boundaries seen in Fig.~\ref{fig:phasediag}. As $k$ is tuned from negative to positive values, the systems display sequence of phase-separated to homogeneous to phase-separated states, corresponding respectively to the chemorepulsive, weakly interacting to chemoattractive regimes. At the intermediate regime, near $k=0$, chemophoretic interactions are too weak to lead to any aggregation. While this qualitative feature bears resemblance to coexistence curves in polymer mixtures having both upper and lower critical solution temperatures~\cite{Patterson1972,SAEKI1976,Praun1991,heikki2017,Blair2024}, the underlying mechanism here is fundamentally different. The re-entrant behavior observed here arises from chemophoretically induced nonreciprocal interactions that dynamically generate correlations in the system. Furthermore, though the coexistence curves bear resemblance to those of an equilibrium vapor-liquid transition, the underlying phases here are intrinsically nonequilibrium. In addition, the extents and locations of phase coexistence depend explicitly on the system's overall packing fraction and composition. This bears, thus, resemblance with those for Vicsek-type active systems~\cite{vicsek1995,Czirok1997,Albano2008}, in which coexistence and critical point are density-dependent rather than universal. 

Building upon our analysis of the chemoattractive case in an earlier work~\cite{mj2025}, we provide here a general picture by including the chemorepulsive case. We
carry out a systematic comparison between the two cases. While chemoattraction promotes clustering through effective attractive interactions between dissimilar particles, chemorepulsion gives rise to aggregation through a fundamentally different mechanism. In the chemorepulsive regime phoretic particles experience activity-induced kicks from the surrounding source particles. Though, this push leads to aggregation, favoring reduction of energy, the mechanism produces strong dynamical trapping. The interplay between aggregation and caging in fact suppresses coarsening. Nevertheless, this overall mechanism stabilizes microphase-separated steady states characterized by finite-sized, fractal-like domains. In contrast, the chemoattractive case is free of any such self-limiting mechanism. In this case, clusters grow uninterrupted, leading to macrophase separation.

\begin{figure}[ht!]
    \centering
    \includegraphics[width=\linewidth]{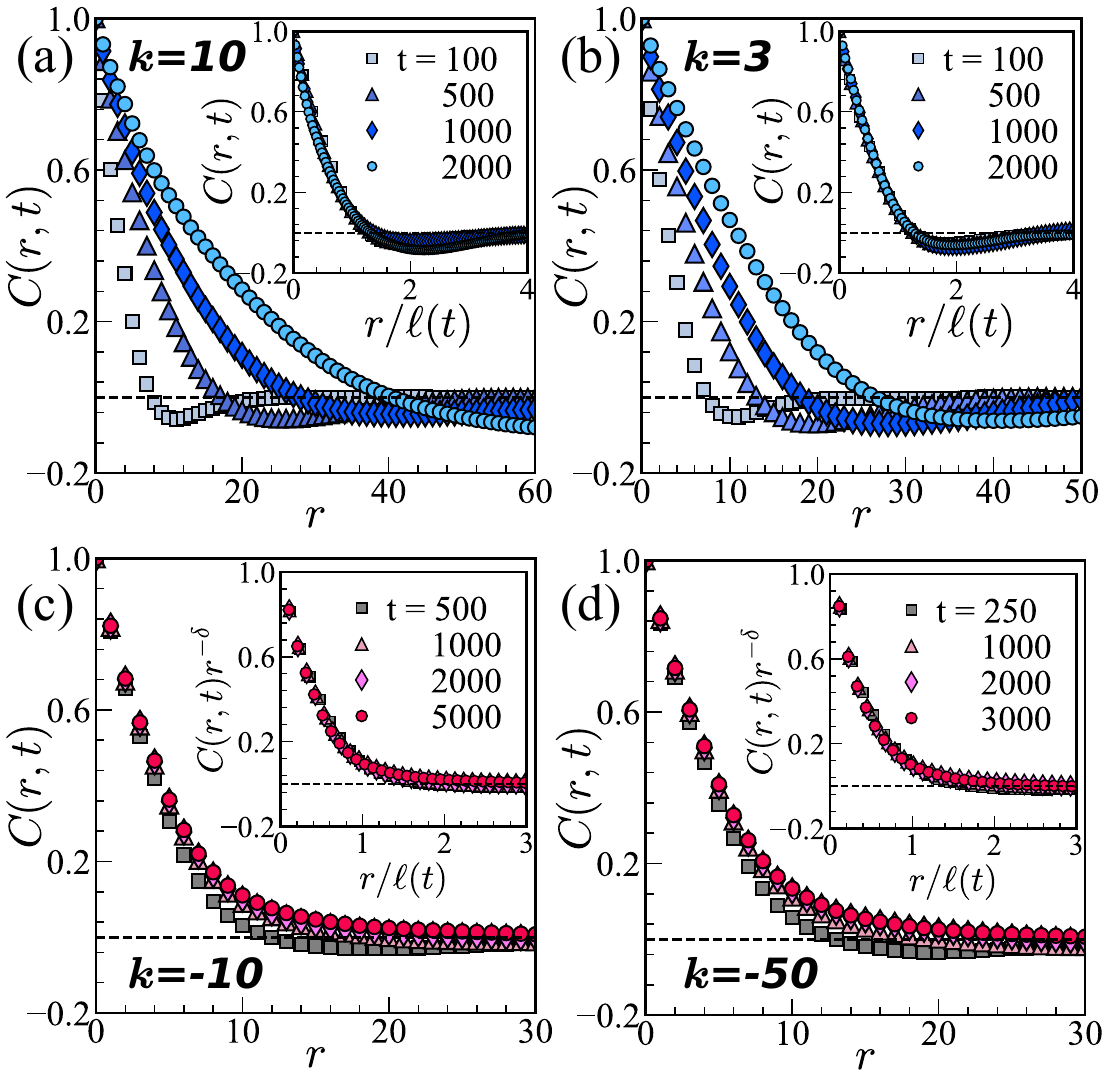}
    \caption{Two-point equal time correlation function, $C(r,t)$, is plotted against $r$. In (a) and (b) we show results for the chemoattractive case with $k=10$ and $k=3$, respectively. The insets of these figures show the scaling plots of $C(r,t)$, where the distance is divided by the time-dependent domain lengths $\ell(t)$. In (c) and (d), we show $C(r,t)$ for the chemorepulsive case with $k=-10$ and $k=-50$, respectively. Due to the fractal nature of clusters in this system, the scaling form of $C(r,t)$ gets modified via the introduction of $d_f$. The dashed horizontal lines represent the ordinate value $``0"$. }
    \label{fig:corr_func}
\end{figure}
To have a quantitative understanding of the difference between the two cases, we calculate the two-point equal-time correlation function $C(r,t)$, defined in Eq.~(\ref{eq:1}). In Fig.~\ref{fig:corr_func} (a, b), we plot $C(r,t)$ for the chemoattractive case, for phoretic strengths $k=10$ and $k=3$, respectively. Fig.~\ref{fig:corr_func} (c) and (d) show the same for the chemorepulsive case, with $k=-10$ and $k=-50$, respectively. Slower decay of the correlation function with the progress of time $t$ indicates growth of domains. Comparing Fig.~\ref{fig:corr_func} (a, b) with (c, d) it is clear that the growth in chemorepulsive case is much slower. Before quantifying the growth, we verify the scaling properties discussed in Eq.~(\ref{eq:scaling}) (see the insets). For this purpose, we divide the distance variable by the characteristic length scale $\ell(t)$ of the system. The latter is obtained by finding the value of $r$ at which $C(r,t)$ decays~\cite{mj2025} to $0.1$. As can be seen in the insets for both Fig.~\ref{fig:corr_func}(a) and (b), the curves at different times collapse, implying self-similar growth of the domains in the chemoattractive case. However, for the chemorepulsive case, the data collapse in $C(r,t)$ did not appear acceptable due to the fractal nature of the clusters. In this case, we carried out a somewhat different scaling exercise~\cite{s1992fractal}:
\begin{equation}{\label{eq:df_scaling}}
    C(r,t)\equiv r^{\delta}\tilde{C}(r/\ell(t)),
\end{equation}
where $\delta=d-d_f$ with $d$ being the spatial dimension. The fractal dimension ($d_f$) is calculated using the box-counting technique~\cite{Mandelbrot1967,s1992fractal,bray2002,LIEBOVITCH1989386}. In this method, the simulation box is divided into smaller boxes of side $a$. The number of boxes, $N(a)$, required to cover the fractal structure, is determined. The fractal dimension is then given by
\begin{equation}{\label{df}}
d_f = \lim_{a \rightarrow 0}\frac{\log N (a)}{\log (1/a)}.\\
\end{equation}
\begin{table}[ht!]
    \centering
    \caption{Dependence of the fractal dimension $d_f$ on $k$.}
    \label{tab:df_k_table}
    \setlength{\tabcolsep}{8pt} 
    \renewcommand{\arraystretch}{1.2}
    \begin{tabular}{lcccccc}
        \toprule
        $k$ 
        & $-2$ 
        & $-10$ 
        & $-20$ 
        & $-30$ 
        & $-40$ 
        & $-50$ \\
        \midrule
        $d_f$ 
        & $2.0$ 
        & 1.93 
        & 1.86 
        & 1.83 
        & 1.80 
        & 1.76 \\
        \bottomrule
    \end{tabular}
\end{table}
As shown in Table~\ref{tab:df_k_table}, $d_f$ decreases from $2.0$ to $1.76$ as $k$ varies from $-2$ to $-50$. This clearly indicates a transition from nearly circular-shaped clusters to extended fractal-like domains as $k$ decreases. By using the scaling form in Eq.~(\ref{eq:df_scaling}) and the relevant number for $d_f$, we obtain high-quality data collapse as shown in the insets of Fig.~\ref{fig:corr_func} (c) and (d).

\begin{figure}[ht!]
\includegraphics[width=\linewidth]{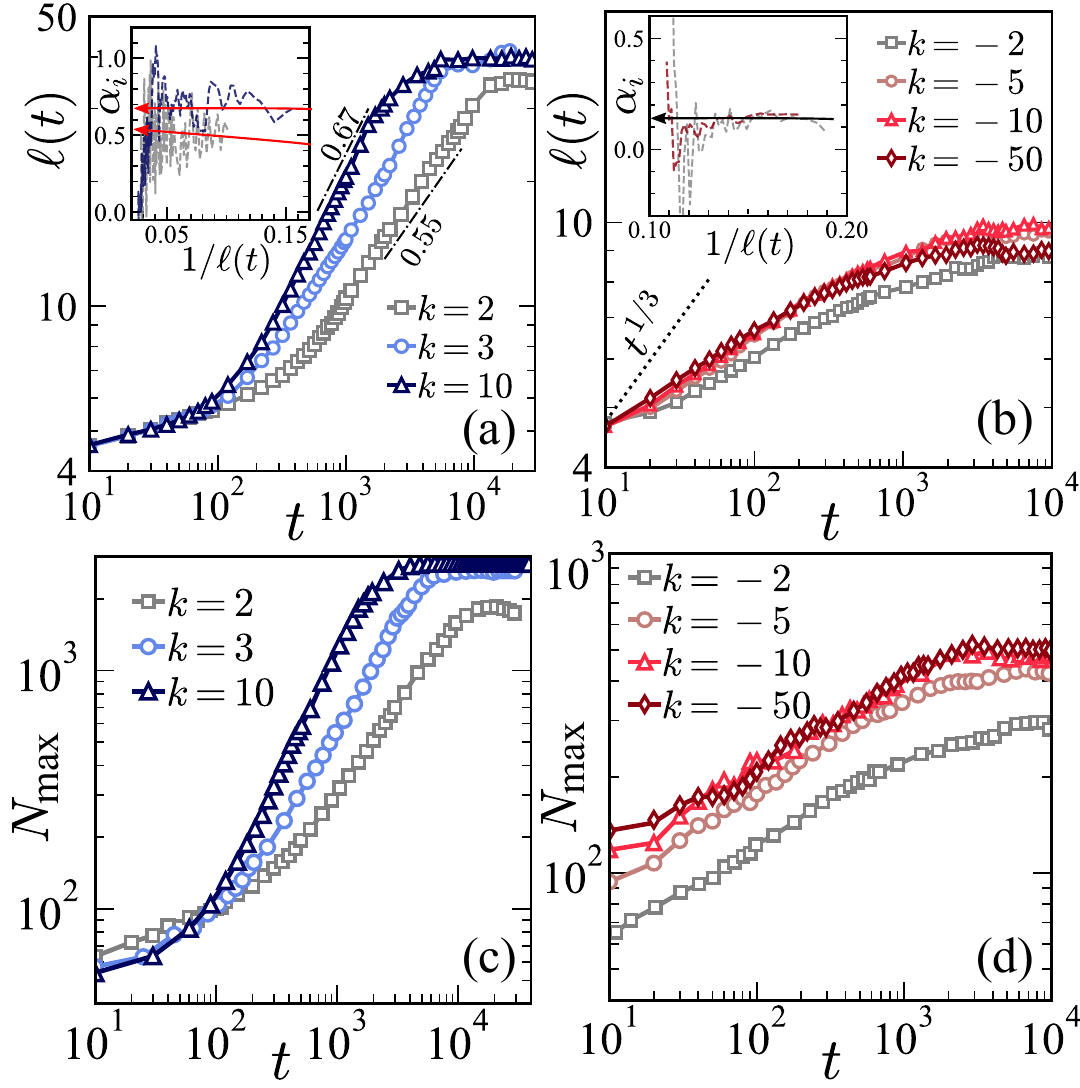}
\caption{Evolutions of domain lengths $\ell(t)$ for different $k$ values, for the cases of (a) chemoattraction and (b) chemorepulsion. These results are obtained from the scaling of $C(r,t)$. The insets of both (a) and (b) show the instantaneous growth exponent, $\alpha_i$, as a function of $1/\ell(t)$. The arrow-headed lines point towards the $\ell \rightarrow \infty$ limit. (c-d) Evolutions of maximum cluster size ($N_{\text{max}}$), as a function of time, for chemoattractive and chemorepulsive cases, respectively.}
\label{fig:domain_len}
\end{figure}

The domain lengths, $\ell(t)$, calculated via the relevant scaling forms of $C(r,t)$, are plotted for the chemoattractive and chemorepulsive cases in Fig.~\ref{fig:domain_len}(a) and Fig.~\ref{fig:domain_len}(b), respectively. It is evident that the domains keep growing, for the chemoattractive case, until hit by the finiteness of the systems. The data on a log-log scale show power-law character with strong dependence of the exponent on $k$. In order to estimate $\alpha$ accurately we calculate the instantaneous exponent~\cite{huse1986corrections} $\alpha_i$ as  
\begin{equation}\label{eq:11}
  \alpha_i = \frac{d \ln \ell}{d \ln t}.
\end{equation} 
The inset of Fig.~\ref{fig:domain_len}(a) shows $\alpha_i$ as a function of $1/\ell(t)$ for $k=2$ and $k=10$. The falls of $\alpha_i$ in the large $\ell$ limit imply finite-size effects. The arrow-headed lines point towards the extrapolated values of $\alpha_i$ in the $\ell \rightarrow \infty$ limit. The numbers are $0.55$ and $0.67$ for $k=2$ and $k=10$ cases, respectively, confirming the above-mentioned strong dependence. 

The evolution in the chemorepulsive case is markedly suppressed, the dynamics being severely constrained. Repulsive interactions, originating from the source particles, trap the phoretic particles, thereby leading to the formation of caged clusters, made purely of phoretic particles. As a result, the characteristic domain size remains significantly smaller than the lateral system dimension, even after prolonged period. As shown in the inset of Fig.~\ref{fig:domain_len}(b), the extracted growth exponent lies well below $1/3$, a value for the slow growth related to Lifshitz-Slyozov diffusive coarsening. Furthermore, in this case, $\alpha$ is only weakly dependent upon the strength of the phoretic interactions as shown in the inset of Fig.~\ref{fig:domain_len}(b). To further quantify the growth, we show the evolution of the size of the largest cluster ($N_{\text{max}}$) in Fig.~\ref{fig:domain_len}(c) and (d) for chemoattraction and chemorepulsion, respectively. It is evident that for chemoattraction the cluster size can grow to large values. On the other hand, chemorepulsion leads to formation of much smaller clusters that do not coarsen continuously to make a single cluster. The $N_{\text{max}}$ behavior further supports the arrested coarsening as suggested by the analysis of $\ell(t)$. 

\section{Mechanism of Growth}

\begin{figure*}
    \includegraphics[width=\linewidth]{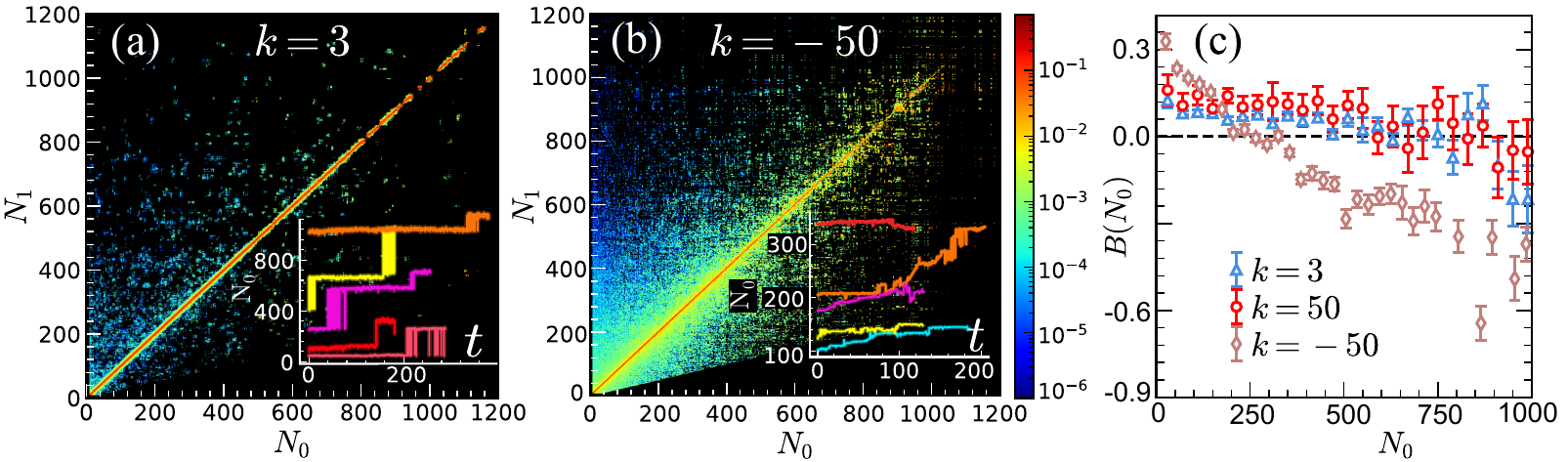}
    \caption{Transition probability matrix $P(N_1|N_0,\tau_s)$, analyzed for time interval $\tau_s=10 \tau$, for (a) chemoattractive $(k=3)$ and (b) chemorepulsive $(k=-50)$ cases. The color-bar represents $P(N_1|N_0,\tau_s)$ on log scale. Insets of (a) and (b) show the number of particles in clusters of different sizes tracked over time $t$. (c) Conditional growth-fragmentation bias $B(N_0)$, for $k=3,~50$  and $k=-50$, are shown against $N_0$. Here $B>0$ indicates growth dominated dynamics and $B<0$ means fragmentation-dominated dynamics.} 
    \label{fig:transition_matrix}
\end{figure*}

The nontrivial values of the growth exponents discussed above do not conform to any known equilibrium coarsening laws. To understand the underlying mechanism, we analyze the dynamics at microscopic level. For this purpose we construct a transition matrix, $P(N_1|N_0,\tau_s)$, related results for which are shown in Fig.~\ref{fig:transition_matrix}. Each matrix element at coordinates $(N_0,N_1)$ represents the probability~\cite{ginot2018aggregation} that a cluster of size $N_0$ at time $t$ evolves into a cluster of size $N_1$ after a time increment $\tau_s$. Diagonal elements therefore quantify cluster persistence. Elements above the diagonal correspond to aggregation events, whereas elements below the diagonal capture fragmentation processes. The color scale represents the probability density $P(N_1|N_0,\tau_s)$ on a logarithmic scale. We have verified that the qualitative features of the transition matrices remain robust over the range $\tau_s \in [0.5 \tau,10\tau]$. Note that the presented results correspond to $\tau_s = 10\tau$. Fig.~\ref{fig:transition_matrix}(a) and (b) display the transition matrices for the chemoattractive ($k=3$) and chemorepulsive ($k=-50$) cases, respectively.

In each of the cases, the transition matrix is sharply peaked along the diagonal, indicating that most clusters retain their sizes over the observation time window. However, pronounced differences emerge in the off-diagonal structure. In the chemoattractive case, the diagonal band is much narrower, and off-diagonal parts contain intermittent and asymmetric probability weights, reflecting a net bias toward cluster-cluster aggregation than fragmentation. These persistent aggregation events drive sustained growth and ultimately lead to macrophase separation. In contrast, the chemorepulsive case exhibits a significantly broader diagonal band. Symmetric nature there is indicative of frequent and reversible particle attachment and detachment. Substantial probability weight appears on both sides of the diagonal, finally revealing that aggregation and fragmentation events occur with comparable likelihood, the frequency remaining high even for larger clusters, implying growth becoming counteracted by fragmentation. Consequently, association-dissociation dynamics in the chemorepulsive regime are far more dynamic than in the chemoattractive case. This, combined with the arrested feature, leads to stabilization of microphase-separated steady states.

To further elucidate the origin of the faster growth observed in the chemoattractive regime, we track the time evolution of individual cluster sizes. Representative individual cluster sizes are shown in the inset of Fig.~\ref{fig:transition_matrix}(a), as a function of time. These trajectories reveal existence of long-lived clusters with sudden, discontinuous increase in size, characteristic of cluster-cluster coalescence events. In contrast, the corresponding results in the chemorepulsive case, shown in the inset of Fig.~\ref{fig:transition_matrix}(b), display a much more gradual and continuous increase in cluster size, consistent with growth via single-particle aggregation rather than cluster coalescence. Importantly, these incremental growth events are frequently followed by fragmentation, resulting in strongly fluctuating cluster sizes. Therefore, transient growth, frequently counteracted by fragmentation, provides a microscopic explanation for the slow and arrested coarsening observed in the chemorepulsive regime.

We isolate the growth and fragmentation events to compute below the conditional growth-fragmentation bias by excluding size-preserving transitions. This quantifies size changes toward aggregation or fragmentation.
We define the bias, for a cluster of size $N_0$, as 
\begin{equation}
    B(N_0) = \frac{P_{+}(N_0) - P_{-}(N_0)}{P_{+}(N_0) + P_{-}(N_0)},
\end{equation}
where
\begin{align}
P_{+}(N_0) &= \sum_{N_1>N_0} P(N_1|N_0,\tau_s), \\
\intertext{and}
P_{-}(N_0) &= \sum_{N_1<N_0} P(N_1|N_0,\tau_s).
\end{align}
Hence, $B(N_0)>0$ indicates a growth dominated regime and $B(N_0)<0$ a fragmentation dominated regime, $B(N_0)=0$ providing information on no net growth regime. $B(N_0)$ for chemoattraction ($k=3$ and $50$) and chemorepulsion ($k=-50$) cases are shown in Fig.~\ref{fig:transition_matrix}(c). The plot reveals a clear quantitative distinction between the two cases. In the chemoattractive case, the bias remains positive over a broad range of cluster sizes. This indicates that when size-changing events occur aggregation is statistically favored. In contrast, the chemorepulsive system exhibits a size-dependent bias: while small clusters show a tendency to grow, larger clusters preferentially fragment, leading to self-limiting growth. In summary, the combined analyses of transition matrices, individual cluster-size evolutions and conditional growth-fragmentation bias reveal that coarsening in the chemoattractive systems is driven by intermittent but persistent cluster aggregation events that leads to macro-phase separation. On the other hand, in the chemorepulsive case growth is transient, owing to frequent fragmentation. This leads to micro-phase separation. This mechanistic distinction explains the markedly different coarsening behavior observed in the two cases. 
  
\begin{figure}
    \centering
    \noindent
\includegraphics[width=0.5\linewidth]{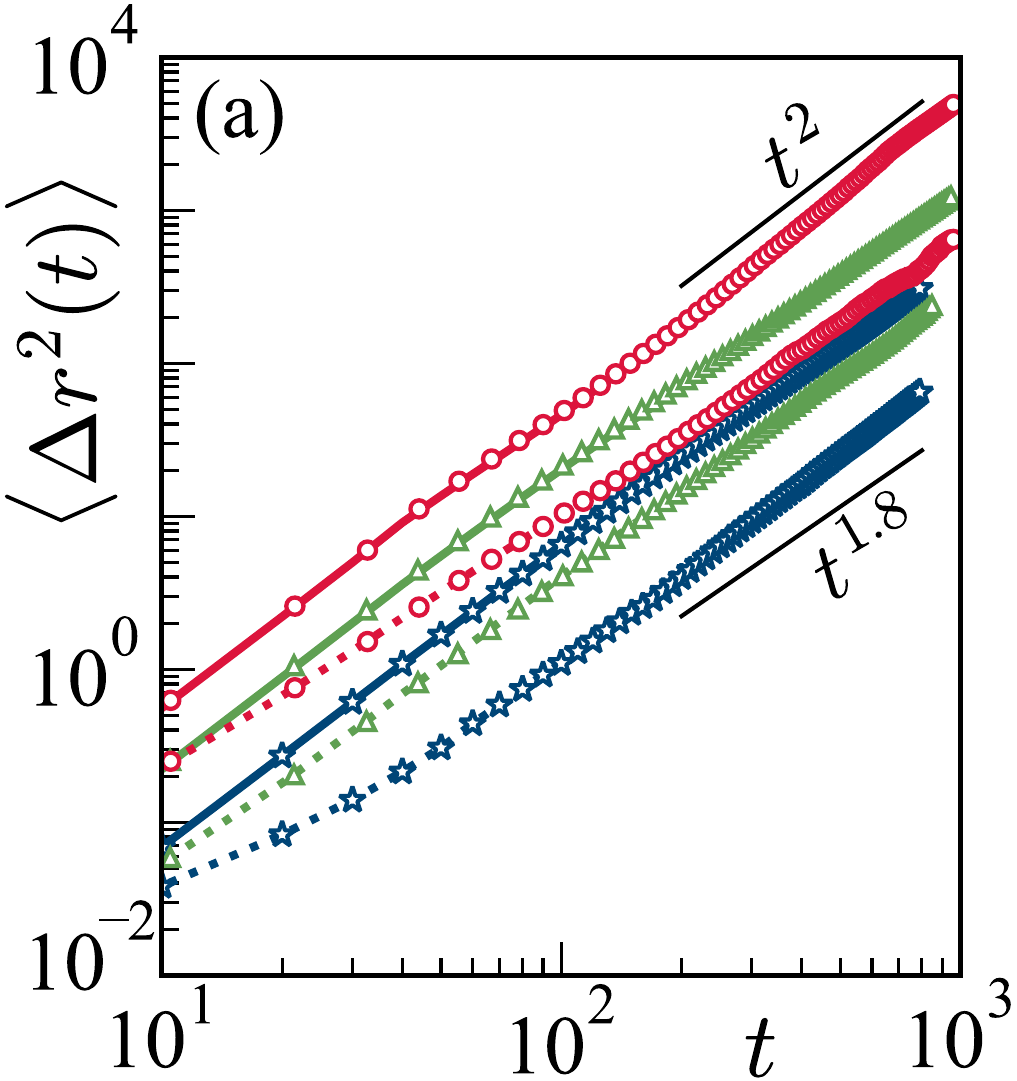}\includegraphics[width=0.5\linewidth]{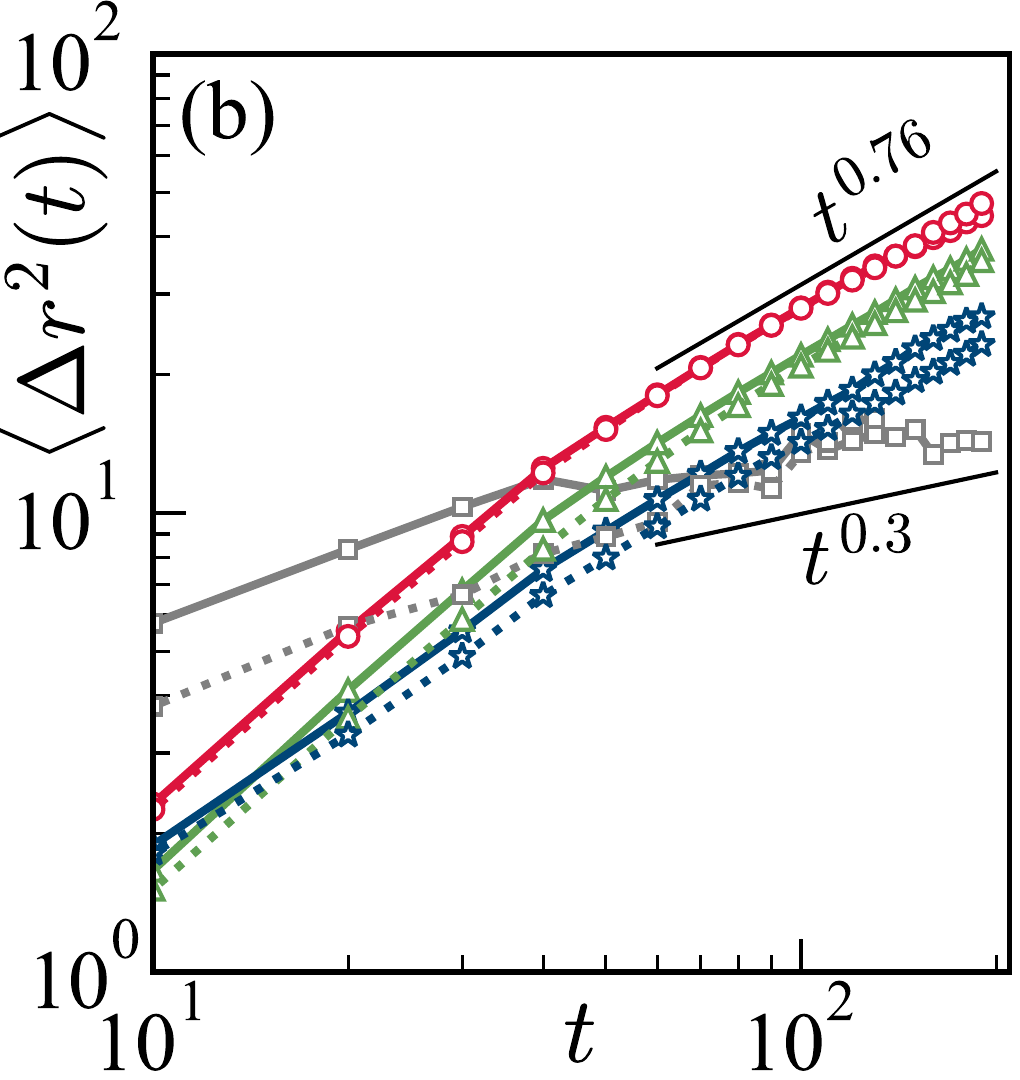}
    \caption{(a) Mean squared displacements of clusters of different sizes in the case of chemoattraction, for two different phoretic strengths $k=10$ (solid lines) and $k=3$ (dotted lines). Different colors correspond to clusters of varying initial sizes: red ($N_C\simeq 150$), green ($N_C\simeq800$) and blue ($N_C\simeq2150$). (b) Same as (a) but for the chemorepulsion case, with phoretic strengths $k=-50$ (solid lines) and $k=-10$ (dotted lines). The colors here denote clusters of following sizes: red ($N_C\simeq50$), green ($N_C\simeq100$), blue ($N_C\simeq250$), and grey ($N_C\simeq750$).}
    \label{fig:msd_cluster}
\end{figure}
To analyze the growth mechanism further, we probe the dynamics of individual clusters via the calculation of mean-squared displacements (MSD), $\langle \Delta r^2(t) \rangle
= \left\langle(\mathbf{r}_{\mathrm{cm}}(t) - \mathbf{r}_{\mathrm{cm}}(0))^2 \right\rangle$, where the position vector of the center of mass of a cluster, having $N_C$ particles, is defined as $\mathbf{r}_{\mathrm{cm}}(t) = (1/N_C)\sum_{i=1}^{N_C} {{\bf{r}}_{i}} (t)$. The cluster MSD for differently sized clusters, for both chemoattraction and chemorepulsion, are presented in Fig.~\ref{fig:msd_cluster}(a) and (b). Typically, at late times $\langle \Delta r^2(t) \rangle \sim  t^{\lambda}$: $\lambda<1$ represents sub-diffusive motion, for $\lambda = 1$ the motion is diffusive, and one has $\lambda = 2$ for ballistic motion~\cite{hansen2013theory}. In Fig.~\ref{fig:msd_cluster} we show MSD for clusters of different sizes. For the chemoattractive case (see part (a)), the clusters exhibit strong superdiffusive to even ballistic behavior that has been correlated with growth exponents $\alpha>1/2$ in an earlier work~\cite{mj2025}. It has been argued that none of the simple limiting pictures, viz., diffusive coalescence or ballistic aggregation~\cite{carnevale1990statistics, trizac2003correlations, trizac1996dynamics}, applies in the chemoattractive coarsening. Rather, the growth here is due to complex superdiffusive motion of the clusters. This is evident from Fig.~\ref{fig:msd_cluster}(a). In contrast, clusters of phoretic particles in the chemorepulsive case display strong sub-diffusive behavior as evident from Fig.~\ref{fig:msd_cluster}(b). The motion is restricted by the surrounding source particles that repel them. The small values of $\alpha$, therefore, are due to slow deposition of other phoretic particles 
on existing clusters.

\section{Caging Dynamics}
\begin{figure}[ht!]
    \centering
        \includegraphics[width=0.9\linewidth]{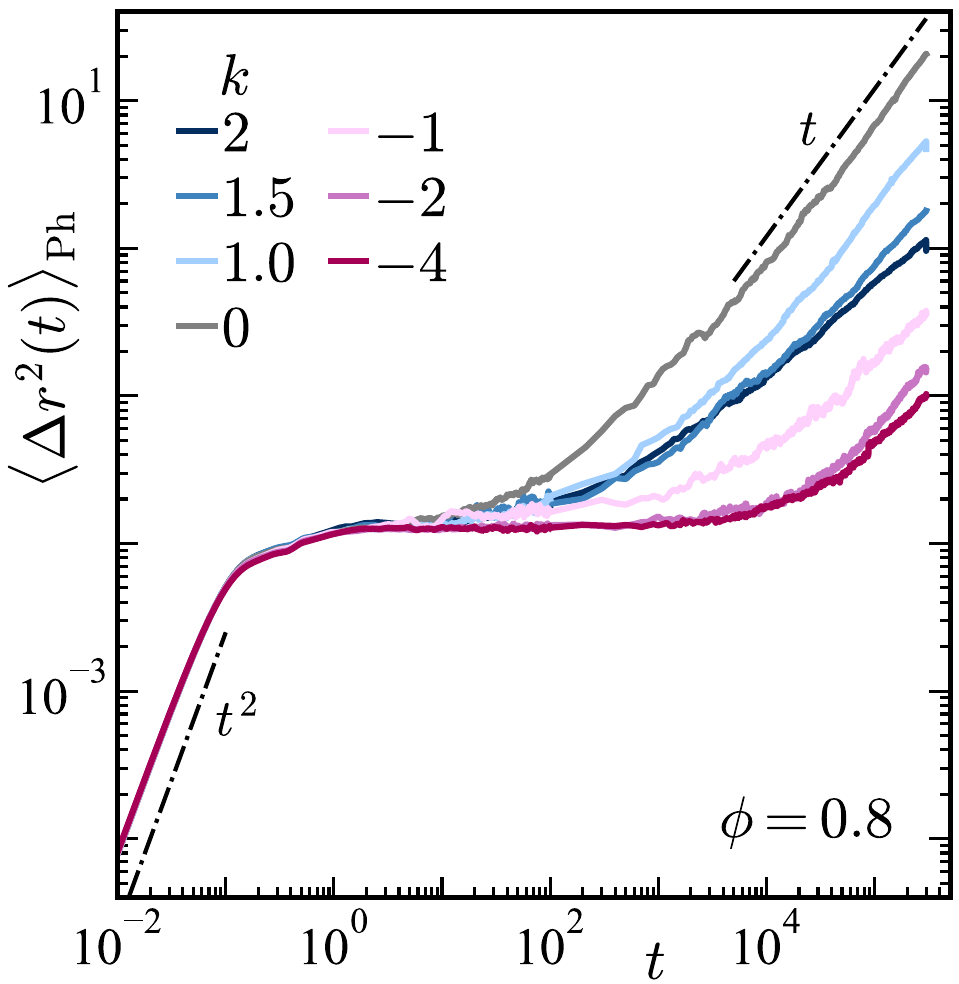}
        \caption{Average mean squared displacements of phoretic particles ($\langle \Delta r^{2}(t) \rangle_{\mathrm{Ph}}$) for varying phoretic strengths. These results are for an overall packing fraction value $\phi=0.8$.}
    \label{fig:msd_glassy}
\end{figure}

We have observed pronounced suppression of phoretic cluster mobility in the chemorepulsive regime. Here we explore the dynamics at single particle level. For this purpose we consider steady-state with $\phi =0.8$ and $T=0.4$, thereby remaining in a regime sufficiently far from the high-density branch of the coexistence curve shown in Fig.~\ref{fig:phasediag}. Staying at this density, we systematically vary the interaction strength $k$ from negative to positive values. This protocol allows us to conveniently probe the role of chemophoretic interactions on particle-level dynamics~\cite{guo2025nonmonotonic}.

Fig.~\ref{fig:msd_glassy} shows the time evolution of the MSD, $\langle \Delta r^{2}(t) \rangle_{\mathrm{Ph}}$, of phoretic particles, spanning both chemoattractive ($k>0$) and chemorepulsive ($k<0$) regimes. For all values of k, the MSD exhibits a short-time ballistic regime. At intermediate and long times, however, the dynamics display strong dependence on $k$. As $|k|$ is increased, the MSD gets progressively suppressed, signaling an increasing degree of dynamical constraint. At long times, particles eventually escape the cages, restoring nearly diffusive transport. In the chemorepulsive regime, even in the long-time limit, the MSD shows a pronounced sub-diffusive behavior over extended time scales, indicative of strong caging effects. This suppression of transport arises from repeated interactions with surrounding source particles, which dynamically confine the phoretic particles and hinder their ability to explore space freely. We note that $\langle \Delta r^{2}(t) \rangle_{\mathrm{Ph}}$ is computed after subtracting the center-of-mass drift of the system from individual particle trajectories.

For weaker interactions, viz., near $k= 0$, the MSD crosses over to a diffusive regime at long times, consistent with relatively unconstrained motion even in a dense but homogeneous environment~\cite{Berthier2011}. Interestingly, even in the chemoattractive regime, the long-time MSD remains suppressed at large $|k|$, reflecting a degree of coherent motion of particles. Overall, the MSD behavior demonstrates that increasing chemophoretic interaction strength, particularly in the chemorepulsive regime, leads to a strong reduction in particle mobility due to dynamical trapping. These single-particle dynamics provide microscopic evidence for the arrested coarsening and microphase-separated steady states observed at the collective level.

\section{Conclusion}\label{sec:conc}

We have investigated the phase behavior, coarsening dynamics, and transport in chemophoretic binary mixtures by systematically varying the interaction strength $k$ and density $\phi$. The steady-state phase diagram reveals interesting sequences of phases as $k$ is tuned from  positive to negative values, corresponding respectively to chemoattractive, weakly interacting, and chemorepulsive regimes. While chemoattraction drives macrophase separation through cluster coalescence, chemorepulsion leads to aggregation via persistent nonequilibrium kicks from source particles. This, however, leads to dynamical trapping and, thus, microphase separation, characterized by finite-sized, fractal-like domains. This  distinction is reflected in quantitative features of the coarsening kinetics. The chemoattractive regime has been shown to exhibit unbounded growth with exponent that depends on $k$. In the chemorepulsive regime, the exponent is largely independent of $k$.

Complementary analysis of individual cluster-size evolution and transition matrices in the two cases also reveals pronounced differences. For the case of chemorepulsion, while small clusters exhibit a tendency to grow, larger clusters preferentially fragment, resulting in self-limiting growth. This dynamical balance between association and fragmentation stabilizes microphase-separated steady states composed of finite-sized domains. At high densities, measurements of mean-squared displacements reveal a crossover from ballistic to strongly sub-diffusive dynamics with increasing $|k|$, providing microscopic evidence of caging and dynamical confinement of phoretic particles, particularly in the chemorepulsive regime. These results highlight an interesting route to re-entrant phase separation and microphase formation that lies outside known equilibrium universality.

\begin{acknowledgments}
MJ acknowledges CSIR India for research fellowship. The authors would like
to acknowledge the HPC facility at IISER Bhopal, India and National Supercomputing Mission (NSM) facility PARAM Shivay at IIT (BHU) Varanasi for computational time. ST acknowledges SERB India for funding via Grant No. CRG/2022/003778.
\end{acknowledgments}

\bibliography{bib_pre}

\end{document}